\documentclass[prb,showpacs,floatfix]{revtex4}
\usepackage{version}
\usepackage{times}
\usepackage{graphics}
\usepackage{color}
\usepackage{amssymb}
\usepackage{bm} 
\newcommand{\p}{\\[2ex]}
\newlength{\figwidth}
\newcommand{\tabl}[1]{table~\ref{#1}}

\newlength{\hw}
\newlength{\vvp}
\newlength{\minusspace}
\newcommand{\msp}{\hspace{\minusspace}}
\newlength{\negspace}
\newlength{\zerospace}
\newcommand{\tstrut}{\rule[-1.8ex]{0ex}{3.5ex}}
\newcommand{\refjl}[5]{#1 (#5) {\it#2} {\bf #3} #4.}
\newcommand{\ymno}{YMn$_2$O$_5$}
\newcommand{\Qv}{\ensuremath{{\bf M}_{\perp}}}
\newcommand{\Qvs}{\ensuremath{{\bf M}_{\perp}^*}}
\newcommand{\rv}{\ensuremath{{\bf r}}}
\newcommand{\lv}{\ensuremath{{\bm \ell}}}
\newcommand{\kv}{\ensuremath{{\bf k}}}

\newcommand{\gv}{\ensuremath{{\bf g}}}
\newcommand{\Pv}{\ensuremath{{\bf P}}}
\newcommand{\Mv}{\ensuremath{{\bf M}}}

\newcommand{\tv}{\ensuremath{{\bf t}}}
\newcommand{\zv}{\ensuremath{{\bf z}}}
		\newcommand{\delv}{\ensuremath{{\bm \delta}}}
\newcommand{\hf}{\ensuremath{\frac12}}
\newcommand{\qr}{\ensuremath{\frac14}}
 \newcommand{\mntp}{Mn$^{3+}$}
 \newcommand{\mnfp}{Mn$^{4+}$}
\newcommand{\pv}{\ensuremath{{\bm \tau}}}
\begin{document}
\title[]{Polarisation dependence of magnetic Bragg scattering in \ymno
}
\author { P J Brown and T Chatterji}
\address{Institut Laue Langevin, BP 156, 38042 Grenoble, France}
\begin{abstract}
The polarisation dependence of the intensity of elastic magnetic scattering from \ymno\ single crystals
has been measured at 25~K in magnetic fields between 1 and 9~T. A significant polarisation dependence was observed in the intensities of magnetic satellite reflections, propagation vector $
\pv=0.5,0,0.25$ measured with both the [100] and [010] axes parallel to the common polarisation and applied field direction. The intensity asymmetries $A$ observed in sets of orthorhombicly equivalent reflections show systematic relationships which allow the phase relationship between different components of their magnetic interaction vectors to be determined. They fix the orientation relationships between the small $y$ and $z$ moments on the \mnfp\ and \mntp\ sub-lattices and lend support to the structure reported by Kim et al. \cite{kim:08}. It was found that that $A(hkl)\ne A(\bar h\bar k\bar l)$ which suggests that there is a small modulation of the nuclear structure which has the same wave-vector as the magnetic modulation leading to a small nuclear  structure factor for the satellite reflections. The differences $A(hkl)- A(\bar h\bar k\bar l)$ observed indicate shifts
in the atomic positions of order 0.005 \AA.
\end{abstract}
\maketitle

\section {Introduction}
\ymno\ together with other members of the series of   isostructural rare
earth compounds (RMn$_2$O$_5$) exhibits
multiferroic properties \cite{kag:02,ino:96,hur:04} of a type unique
amongst known multiferroics. Whereas the magnetic structures giving rise to ferroelectricity in other materials are some form of
spiral (cycloidal, conical etc.) the magnetic order in the ferroelectric phases of RMn$_2$O$_5$ compounds has been shown to be only very weakly cycloidal, with the major components of the moments arranged nearly co-linearly in the $x$-$y$ plane and modulated as a spin density wave. The electric polarisation and several other
observations made on these phases can be explained within the framework of a simple, symmetric
exchange
model, in which polarisation is generated by superexchange striction,
effective even when the spins are perfectly collinear
\cite{chap:04,blake:05,chap:06}.
This is in sharp contrast to the mechanism proposed for the
RMnO$_3$ series and other multiferroics in which  non-collinear spins are  needed not
only to break inversion symmetry, but also to generate atomic displacements,
through the
inverse Dzyaloshinsky-Moriya effect \cite{serg:06,sawa:05}.\\[1ex]
The present experiment was undertaken to study the effect of external magnetic fields on the population of both chiral and ferroelectric domains in \ymno\ and  to determine  whether information about the atomic displacements leading to ferro-electricity can be obtained from measurement of the polarisation dependence of the intensities of the magnetic reflections,
\p
{\bf  Polarisation dependence of the intensity of magnetic satellite reflections}\\
The cross-section for elastic neutron scattering of a beam with polarisation \Pv\ from a magnetic crystal is given by
\begin{equation}
I^{\Pv} \propto NN^* + \Qv\cdot\Qvs + 2\Re (N\Pv\cdot\Qvs)+ \Pv\cdot\imath(\Qv\times\Qvs)
\label{polxsec}
\end{equation}
 The purely magnetic intensity scattered by the satellite reflections can be polarisation dependent if the final
 term in eqn~\ref{polxsec} is non-zero. This can only occur if
 if the magnetic structure lacks a centre of symmetry and the moments are non-collinear ($\Qv \nparallel\Qvs$), the populations of the two centro-symmetrically related domains are unequal and
  the scattering vector is not perpendicular to the polarisation direction.
The third term in eqn~\ref{polxsec} can be non-zero only if
the nuclear structure is modulated with the same wave-vector as the magnetic one so that there is a finite  nuclear contribution
$N$ to the magnetic reflections.
\p
The polarised neutron intensity asymmetry $A$ is defined as
$A=(I^+-I^-)/(I^++I^-)$ where $I^+$ and $I^-$ are the intensities scattered
with neutrons polarised respectively parallel and antiparallel to the polarisation direction \Pv.
The intensity asymmetries due to purely magnetic scattering have different symmetry properties from those due to nuclear magnetic interference. Both $\Qv(\kv)$ and $N(\kv)$ obey Friedel's law so
\[\Qv(\kv)=\Qv(\bar\kv)^*\quad\mbox{and}\quad N(\kv)=N(\bar\kv)^* \] so that on reversal of the scattering vector $\Qv(\kv)\times\Qvs(\kv)$ changes sign whilst $\Re\{N(\kv)\Qvs(\kv)\}$ does not. If $A(\kv)$ and $A(\bar\kv)$ are measured in normal beam geometry with the rotation axis parallel to the polarisation direction then
\[\begin{array}{cccccccc}
\frac{(\Qv(\kv)\times\Qvs(\kv))\cdot\Pv}{|N|^2+|M|^2}&=&A_M(\kv)&\mbox{and}&\frac{2\Re\{N(\kv)\Qvs(\kv)\cdot\Pv\}}{|N|^2+|M|^2}&=&A_N(\kv)&\mbox{and }\tstrut\\ 
\frac{(\Qvs(\kv)\times\Qv(\kv))\cdot\Pv}{|N|^2+|M|^2}&=&A_M(\bar\kv)&\mbox{and}&\frac{2\Re\{N(\kv)^*\Qv(\kv)\cdot\Pv\}}{|N|^2+|M|^2}&=&A_N(\bar\kv)&\mbox {so}\tstrut\\
A_M(\kv)&=&-A_M(\bar\kv)&\mbox{and}&A_N(\kv)&=&A_N(\bar\kv)
\end{array}\]
\section {Experimental}
A single crystal of \ymno\ was mounted in the 10~T vertical field superconducting magnet on the polarised neutron diffractometer D3 at ILL Grenoble with its [100] axis parallel to the field direction. The crystal was first cooled to 25~K in a field
of 5~T and the flipping ratios $R$ of 160 magnetic reflection $h k l$ $h=\pm0.5$ were measured. Where possible all 8
orthorhombically equivalent reflections were included. The intensities of almost all the reflections measured were found
to have a small but significant polarisation dependence. To check whether the polarisation dependence arose solely because the crystal had been cooled through the N\'eel transition in a magnetic field the field was reduced to zero, the crystal warmed to 45~K and then cooled again to 25K.  A sub-set of reflections
was then remeasured first with 1~T and then with 3~T applied. Very little change in the flipping ratios was found.
To be sure that the crystal had indeed been cooled through the N\'eel transition in zero field is was warmed to 75~K
and again cooled to 25~K, both in zero field. Measurement of the same set of reflections in 1~T again gave nearly unchanged results.

\begin{table}[htb]
  \caption{Polarised neutron intensity  asymmetries $A$ measured at 25~K in different fields $H$ after cooling in 5~T and in zero field}
  \setlength{\hw}{-1.5ex}
 \begin{tabular}{ccccccc}
  \hline
\hline&&&\multicolumn{4}{c}{Field $\parallel [100]$}\\[\hw]
$h$&$k$&$l$\\[\hw]
&&&$H=5$ T$^a$&$H=1$ T$^b$&$H=3$ T$^b$&$H=1$ T$^c$\\
\hline
$\msp  0.5$&$  1.0$&$\msp 0.25$&$    -0.033(2)$&$    -0.025(6)$&$    -0.041(2)$&$    -0.038(2)$\\
$\msp  0.5$&$  1.0$&$    -0.25$&$\msp 0.041(2)$&$\msp 0.031(6)$&$\msp 0.045(2)$&$\msp 0.043(2)$\\
$\msp  0.5$&$  2.0$&$\msp 0.75$&$    -0.015(4)$&$    -0.016(4)$&$    -0.031(7)$&$    -0.019(4)$\\
$\msp  0.5$&$  2.0$&$    -0.75$&$\msp 0.014(4)$&$\msp 0.018(4)$&$\msp 0.034(7)$&$\msp 0.020(4)$\\
\hline
 &&&\multicolumn{4}{l}{{\footnotesize$^a$Field cooled from 300~K in 5~T.}}\\
	&&&\multicolumn{4}{l}{\footnotesize{$^b$Zero Field cooled from 45~K.}} \\
	&&&\multicolumn{4}{l}{\footnotesize{$^c$Zero Field cooled from 75~K. }}
 	\end{tabular}\\
 \begin{tabular}{cccccc}\hline
&&&\multicolumn{3}{c}{Field $\parallel [010]$ Zero field cooled}\\[\hw]
$h$&$k$&$l$\\[\hw]
&&&$H=1$ T&$H=3$ T&$H=9 T$ \\
$\msp  1.5$&$  1.0$&$\msp 0.25$&$\msp 0.013(2) $&$\msp 0.011(2) $&$\msp 0.010(2)$\\
$\msp  1.5$&$  1.0$&$    -0.25$&$    -0.010(3) $&$    -0.015(2) $&$    -0.011(2) $\\
$\msp  0.5$&$  1.0$&$\msp 1.25$&$\msp 0.019(5) $&$\msp 0.029(5) $&$\msp 0.032(7)$\\
$\msp  0.5$&$  1.0$&$    -1.25$&$     0.054(10)$&$    -0.050(12)$&$    -0.0345(13)$\\
\hline \hline
 	\end{tabular}
   \label{AvsH}
\end{table}
\begin{table}
\begin{center}
\settowidth{\minusspace}{$-$}
\setlength{\hw}{-1.6ex}
\caption{Intensity asymmetries $A$ measured for  groups of  orthorhombicaly equivalent satellite reflections
from \ymno\ at 25~K with [100] and [010] axes parallel to the polarisation direction}
\vspace{0.5ex}
\begin{tabular}{llllllll}
\hline
\multicolumn{4}{c}{Polarisation $\parallel [100]$}&\multicolumn{4}{c}{Polarisation $\parallel [010]$}\\
\multicolumn{1}{c}{$h$}&\multicolumn{1}{c}{$k$}&\multicolumn{1}{c}{$l$}&\multicolumn{1}{c}{$A(hkl)$}&
\multicolumn{1}{c}{$h$}&\multicolumn{1}{c}{$k$}&\multicolumn{1}{c}{$l$}&\multicolumn{1}{c}{$A(hkl)$}\\
\hline
$\msp  0.50$&$\msp  1.00$&$\msp  0.25$&$     -0.034(2)$&$\msp  1.50$&$\msp  1.00$&$\msp  0.25$&$\msp 0.0190(4)$\\
$\msp  0.50$&$\msp  1.00$&$     -0.25$&$\msp  0.041(2)$&$\msp  1.50$&$\msp  1.00$&$     -0.25$&$     -0.021(4)$\\
$     -0.50$&$\msp  1.00$&$\msp  0.25$&$     -0.029(2)$&$     -1.50$&$\msp  1.00$&$\msp  0.25$&$     -0.016(3)$\\
$     -0.50$&$\msp  1.00$&$     -0.25$&$\msp  0.032(2)$&$     -1.50$&$\msp  1.00$&$     -0.25$&$\msp  0.015(6)$\\
$\msp  0.50$&$     -1.00$&$\msp  0.25$&$     -0.056(2)$\\
$\msp  0.50$&$     -1.00$&$     -0.25$&$\msp  0.054(2)$\\
$     -0.50$&$     -1.00$&$\msp  0.25$&$     -0.079(2)$\\
$     -0.50$&$     -1.00$&$     -0.25$&$\msp  0.076(2)$\\
\\
$\msp  0.50$&$\msp  2.00$&$\msp  0.25$&$     -0.033(2)$&$\msp  0.50$&$\msp  1.00$&$\msp  1.25$&$\msp  0.017(3)$\\
$\msp  0.50$&$\msp  2.00$&$     -0.25$&$\msp  0.033(2)$&$\msp  0.50$&$\msp  1.00$&$     -1.25$&$     -0.026(6)$\\
$     -0.50$&$\msp  2.00$&$\msp  0.25$&$     -0.022(2)$&$     -0.50$&$\msp  1.00$&$\msp  1.25$&$     -0.018(3)$\\
$     -0.50$&$\msp  2.00$&$     -0.25$&$\msp  0.019(2)$&$     -0.50$&$\msp  1.00$&$     -1.25$&$\msp  0.024(6)$\\
$\msp  0.50$&$     -2.00$&$\msp  0.25$&$     -0.038(2)$\\
$\msp  0.50$&$     -2.00$&$     -0.25$&$\msp  0.033(2)$\\
$     -0.50$&$     -2.00$&$\msp  0.25$&$     -0.064(2)$\\
$     -0.50$&$     -2.00$&$     -0.25$&$\msp  0.062(2)$\\
\hline
\end{tabular}
\label{orthequiv}
\end{center}
\end{table}

The crystal was then remounted with a [010] axis parallel to the field direction. It was cooled to 25~K in zero field and the flipping ratios of a group of 42 magnetic reflections $hkl\quad k=0,1,2$ measured with  1, 3 and 9~T applied.  Again significant polarisation dependence was found in the intensities of almost all the reflections with $k\ne0$, but no significant effect due to increasing the field was observed. Table \ref{AvsH} shows the intensity asymmetry $A$ measured with each applied field for two reflections in
each of the two crystal orientations,  it can be noticed that the signs of $A$ for reflections $hkl$ and $hk\bar l$ are always different.
\\[1ex]
\section {Results}
In table \ref{orthequiv} the asymmetries $A$ of groups of orthorhombically equivalent reflections measured with both crystal orientations are listed. Within each group
it can be seen that
\begin{eqnarray*}
A(hkl)&\approx &-A(\bar hkl)\approx -A(hk\bar l)\approx A(\bar hk\bar l)\quad\mbox{for the [100] orientation}\label{xsigns}\\
A(hkl)&\approx &A(\bar hkl)\approx -A(hk\bar l)\approx -A(\bar hk\bar l) \quad\mbox{for the [010] orientation}\label{ysigns}
\end{eqnarray*}
but that for the [100] orientation where the geometry allowed measurement with $h=\pm 0.5$ $A(hkl)\ne A(\bar h\bar k\bar l)$.
These symmetry relationships were found to apply to all the groups of equivalent reflections measured.\\[1ex]
For the data presented in \tabl{orthequiv} not only the signs, but also the magnitudes of $A(hkl)$ and $A(\bar h\bar k\bar l)$ differ. This suggests that $A$ contains contributions from both $A_M$ and $A_N$ but that the major contribution is from $A_M$ which dictates the sign, the difference in magnitude being due to a smaller contribution from $A_N$.
\section {Magnetic scattering from \ymno}

The magnetic structure of the commensurate phase ($\tau_z=0.25$ ) of \ymno\ has been described in terms of
zigzag antiferromagnetic chains of Mn moments lying in the $ab$ plane  with weak z components modulated in phase quadrature giving the structure a cycloidal character \cite{blake:05,vecchini:08}.  In these structure refinements the components of moment on all the Mn sub-lattices were allowed to vary independently, making the symmetry of the structural motif triclinic. A very similar structure which conserves orthorhombic symmetry  has also been reported \cite{kim:08}. This latter structure is defined using a formalism based on the results
of representation theory \cite{harris:07}. The magnetic moment distribution is characterised by complex order parameters, one for each
of the two irreducible representations given by the symmetry analysis, and by a set of complex vectors giving the moments on the independent magnetic sites associated with each of the representations.  With this model the magnetic structure factors can be expressed as 
\begin{equation}
\Mv_{\pv}(\kv)=\sum_{r}\sum_{i}\sum_{j}f_i(k)\sigma_r \tilde {O_{rlj}}\Mv_{ir}\exp(\imath\kv\cdot(\tilde R_j\rv_i+\tv_j)+\phi_{j})
\end{equation}
The subscript $r$ labels the representation, $i$ the independent magnetic sites (Mn$^{3+}$, Mn$^{4+}$) and $j$ the elements of the space group Pbam. The $\tilde R_j:\tv_j$ 
are the real space symmetry operations and $\tilde{O}=\tilde T:\tilde C:\tilde E$ are operators relating the magnetic moment vectors $\Mv_i$ of 
sites generated by these elements, $\tilde T$ is a diagonal rotation matrix, $\tilde C=-1$  a conjugation operation and $\tilde E=-1$ indicates exchange of
the two magnetic moment vectors associated with the site.
$\phi_{ij}$ is a phase factor needed when the symmetry operation generates positions outside the origin cell and the  $f_i(k)$ are the 
appropriate magnetic form factors. The $\tilde{O_j}$ obey the multiplication table of the point group $mmm$. In the structure determined 
by Kim et al. \cite{kim:08} only one of the two possible representations is present and the $\tilde O$ for this representation are given 
in \tabl{symops}.
 With this representation two vectors $\Mv_{il}$ are needed for each set of general equivalent positions but the site symmetries, 
 $m_z$ for Mn$^{3+}$ and $2_z$ for Mn$^{4+}$, limit the number of independent components. For Mn$^{3+}$  element 6 generates the 
same position as element 1 and element 2 the same position as element 5 so $\Mv_{11}=-\rho\Mv_{11}^*$ and $\rho \Mv_{12}=-\Mv_{12}^*$ so the 
$x$ and $y$ components of both $\Mv_{11}$ and $\Mv_{12}$ are real whilst their $z$ components are imaginary. For Mn$^{4+}$ element 1 generates the same position as element 2 so $\Mv_{21}=\rho\Mv_{22}$. In terms of the parameters used by Kim et al. \cite{kim:08}:
\[\Mv_{11}=\rho\rv_2;\ \Mv_{12}=-\rv_1^*\ \mbox{for Mn$^{3+}\quad$and}\quad\Mv_{21}=\xi\zv^*; \Mv_{22}=\tau\zv; \ \mbox{for Mn$^{4+}$}\] 

\begin{table}[htdp]
\caption{Symmetry operators for space group Pbam and the associated operations $\tilde{O_j}$ acting on the magnetic moment vectors for the irreducible representation found by Kim et al. \cite{kim:08} in \ymno. The final columns give the numbers assigned by Harris et al. \cite{harris:08a} and Kim et al. \cite{kim:08} to the positions generated for the 4h (Mn$^{3+}$) and 4f (Mn$^{4+}$) sites. 
}
\begin{center}
\begin{tabular}{llllllccccc}
\multicolumn{3}{c}{Operator$\{\tilde {R_j}:\tv_j\}$}&\multicolumn{4}{c}{Operator $\tilde{O_j}$}&\multicolumn{2}{c}{Mn$^{3+}$}&\multicolumn{2}{c}{Mn$^{4+}$}\\
$j$&&&\multicolumn{1}{r}{$\tilde {T_j}^a$}
&\multicolumn{1}{r}{$\tilde {C_j}$}&\multicolumn{1}{r}{$\tilde{E_j}$}&$\phi_j$&H$^b$&K$^b$&H$^b$&K$^b$\\
1&E&$x,\ y,z$&$\msp\epsilon$&$\msp1$&$\msp1$&$0$&1&1&5&1\\
2&$2_z$&$\bar x,\ \bar y,\ z$&$\msp\rho$&$\msp1$&$-1$&$0$&4&2&5&1\\
3&$2_{1x}$&$\hf+x,\ \hf-y,\ \bar z$&$\msp\tau$&$-1$&$-1$&-$\frac\pi2$&2&4&7&3\\
4&$2_{1y}$&$\hf-x,\ \bar\hf+y,\ \bar z$&$\msp\xi$&$-1$&$\msp1$&-$\frac\pi2$&3&3&7&3\\
5&I&$\bar x,\ \bar y,\ \bar z$&$-\epsilon$&$-1$&$-1$&-$\frac{\pi}2$&4&2&8&4\\
6&$m_z$&$x,\ y,\ \bar z$&$-\rho$&$-1$&$\msp1$&-$\frac\pi2$&1&1&8&4\\
7&$b_x$&$\hf-x,\ \hf+y,\ z$&$-\tau$&$\msp1$&$\msp1$&$0$&3&3&6&2\\
8&$a_y$&$\hf+x,\ \hf-y,\ z$&$-\xi$&$\msp1$&$-1$&$0$&2&4&6&2\\ 
\end{tabular}
\end{center}
{\footnotesize
$^a$ Diagonal rotation matrices with diagonal elements $\epsilon=\ 1, \msp1 ,\msp1;\quad\rho=-1, -1, \msp1;
\hspace*{\zerospace} \quad \tau=\ 1, -1, -1;\quad \xi=-1, \msp1, -1;$\\
$^b$ The numbers given to the Mn ions generated by the operators $j$:  H in reference \cite{harris:08a} and K in reference \cite{kim:08} }
\label{symops}
\end{table}%
\begin{table}[htdp]
\caption{Phase differences between components of the magnetic interaction vectors \Qv\  for orthorhombically related magnetic reflections   structure in a structure with the symmetry of \tabl{symops}.}
\begin{center}
\setlength{\hw}{-1.5ex}
\begin{tabular}{cccccccc}
\hline

Reflection&\multicolumn{1}{c}{$\phi_y-\phi_z$}&\multicolumn{1}{c}{$\phi_z-\phi_x$}
&${S_{yz}}^{\ a}$&${S_{zx}}^{\ a}$&$C_{Nx}\ ^b$\\
\hline
$\ h\  k\  l$          &$\phi_{yz}            $&$\phi_{zx}$        &$+1$&$+1$&+1&\\
$\ h\   k\ \bar l$     &$\phi_{yz}+(2n+1)\pi  $&$\phi_{zx}+(2n+1)\pi$    &$-1$&$-1$&$-1$&\\
$\ \bar h\ k\ l$       &$\phi_{yz}+2n\pi      $&$\phi_{zx}+(2n+1)\pi$    &$+1$&$-1$&$+1$\\
$\ \bar h\  k\ \bar l$ &$\phi_{yz}+(2n+1)\pi  $&$\phi_{zx}+2n\pi$   &$-1$&$+1$&$-1$\\
$\ h\ \bar k\ l$       &$-\phi_{yz}+(2n+1)\pi  $&$-\phi_{zx}+2n\pi$   &$+1$&$ $&$-1$\\
\hline
\end{tabular}\\[0.5ex]
\end{center}
{\footnotesize
$^a$The relative signs observed in the measurements made of $A_M$ and\quad $^b A_N$\\ }
\label{phases}
\end{table}%
To see the symmetry relationships between orthorhombically related magnetic interaction vectors it is convenient to write
\begin{equation} \Qv=\sum_{\alpha=x,y,z}M_\alpha\exp\imath \phi_\alpha\qquad \mbox{then}
\end{equation}
\begin{eqnarray}
 A_M&={}&M_yM_z\sin(\phi_y-\phi_z)|\Qv|^2=M_{yz}S_{yz}\quad\mbox{ for the [100] orientation}
\nonumber\\
 A_M&={}&M_zM_x\sin(\phi_z-\phi_x)|\Qv|^2=M_{zx}S_{zx}\quad\mbox{ for the [010] orientation.}\\
\mbox{and}&  &S_{\alpha\beta}=\sin(\phi_\alpha-\phi_\beta)/|\sin(\phi_\alpha-\phi_\beta)|
\nonumber
\end{eqnarray}
 The relationships between the phases $\phi_\alpha$ for a structure with the symmetry of \tabl{symops} are shown in table~\ref{phases}.
$S_{yz}$ and $S_{zx}$ give the relative signs of the asymmetries observed in symmetrically related
reflections measured with the [100] and [010] orientations. It can be seen that this symmetry predicts the correct relationships for the signs of both $S_{yz}$ and $S_{zx}$ .
\section {Nuclear contribution to magnetic reflections in a modulated magnetic structure }
It has been proposed that ferroelectricity in the {\em commensurate phases} ($\pv=0.5,0,0.25$) of the RMn$_2$O$_5$ series is due to small acentric displacements of the \mntp\ ions   which lower the space group from $Pbam$ to $Pb2_1m$\ \  \cite{chap:04}. If the displacements depend on  both the  magnitude  and the sign of the local  moment  the positions of an atom previously at position \rv\
becomes:\[ \rv + \delv_o \cos(\pv\cdot\lv+\phi_i)+\delv_e \cos^2(\pv\cdot\lv+\phi_i)\] where $\pv$ is the magnetic propagation vector, $\phi_i$ the phase of the modulation for the atom at \rv\ and \lv\  a lattice vector. $\delv_e$ and
$\delv_o$ are the maxima of the even ($\cos^2$) and odd ($\cos$) terms in the modulation of the displacements. The nuclear structure factor can then be written:
\begin{equation}
 N=\sum_{\lv}\sum_{i}b_i\exp\imath\kv\cdot[\lv + \rv_i + \delv_{io} \cos(\pv\cdot\lv+\phi_i)+ \delv_{ie} \cos^2(\pv\cdot\lv+\phi_i)]
\end{equation}
If the displacements are small so that $\kv\cdot\delv<<1$ then $\exp\imath\kv\cdot\delv\approx 1+\imath\kv\cdot\delv$:
\begin{eqnarray}
N&=&\sum_{i}b_i\exp(\kv\cdot\rv_i)\sum_{\lv}\exp\imath\kv\cdot \lv\nonumber\\
&{}&\times \left(1+\imath\kv\cdot\delv_{io}\cos(\pv\cdot\lv+\phi_i)+\imath\kv\cdot\delv_{ie}\cos^2(\pv\cdot\lv+\phi_i)\right)\nonumber\\
\end{eqnarray}
Expressing the cosines in terms of exponentials gives
\begin{eqnarray}
 N =& \sum_{i}&b_i\exp(\kv\cdot\rv_i)\sum_{\lv}\left(\exp\imath\kv\cdot\lv+(\delv_{ie}\exp\imath\kv\cdot\lv)/2\nonumber\right.\\[-1ex] 
&&+(\imath\kv\cdot\delv_{io})\left(\exp\imath[\lv\cdot(\kv+\pv)+\phi_i]+ \exp\imath[\lv\cdot(\kv-\pv)-\phi_i]\right)/2 \nonumber\\
&&\left . +\left(\imath\kv\cdot\delv_{ie}\right)\left(\exp\imath[\lv\cdot(\kv+2\pv)+2\phi_i]+ \exp\imath[\lv\cdot(\kv-2\pv)-2\phi_i]\right)/4\right)
\label{fcal}
\end{eqnarray}
All the terms  except the first in the lattice sum of eqn~\ref{fcal} are due to the displacements, but only the third, which is proportional to
$\delv_{o}$, is non-zero for  the wave-vectors $\kv=\gv\pm\pv$ characterising the anti-ferromagnetic modulation. So only the displacements $\delv_o$ can
give rise to a polarisation dependent cross-section for the antiferromagnetic reflections.
The nuclear structure factor for the {\em magnetic} reflections is then just
\begin{eqnarray}
N_{\tau}=\sum_i\imath b_i(\kv\cdot\delv_{oi})\exp(\kv.\rv_i\pm\phi_i)\\[-1ex]
\quad\mbox{\footnotesize  the $+$ sign for  $\kv=\gv-\pv$ and the $-$ sign for $\kv=\gv+\pv$}\nonumber
\end{eqnarray}
$|\Qv|^2 >> |N_{\tau}|^2$ so that the nuclear contribution to the asymmetry is
\begin{equation}
A_N\approx 2\Re(N_{\tau}\Pv\cdot\Qv)/|\Qv|^2
\label{poleq}
\end{equation}
 In order for nuclear scattering to contribute to the magnetic reflections the nuclear modulation must depend on the sign as well as the magnitude of the magnetic moments. Nuclear displacements, such as those proposed by Chapon et al. \cite{chap:04} which depend only on the magnitude of the moments, lead to finite nuclear scattering for $\kv=\gv$ and $\gv\pm 2\pv$. An interaction between the magnetic polarity of an atom and the electric polarity of its nuclear environment is the fundamental requirement for a magneto-electric effect and will be strongly linked to multiferroic behaviour.
The polarisation dependence of the intensities of the magnetic satellite reflections provides a possible means to study  this effect.

Writing $N_{\tau}=|N_{\tau}|\exp\imath\phi_N$
\begin{eqnarray}
  A_N&\approx&2|N_{\tau}|M_x\cos(\phi_N-\phi_x)/|\Qv|^2=C_{Nx}N_{x} \quad\mbox{for the [100] orientation} \nonumber\\
  A_N&\approx&2|N_{\tau}|M_y\cos(\phi_N-\phi_y)/|\Qv|^2=C_{Ny}N_{y} \quad\mbox{for the [010] orientation}\\
\mbox{and}&  &C_{N\alpha}=\cos(\phi_N-\phi_\alpha)/|\cos(\phi_N-\phi_\alpha)|
\nonumber
\end{eqnarray}
\section{Analysis of the data}
The phase relationships imposed by the symmetry operators in real space (relating different domains), are the same as those imposed by
the equivalent reciprocal space operators (relating equivalent reflections from the same domain) except that the real space operators act also on the polarisation direction, whilst the reciprocal space ones do not.
\p
For the symmetry used here the only domains which can occur are chiral ones corresponding to propagation vectors \pv\  and -\pv. Each asymmetry $A$ measured contains contributions from both chiral domains. If L is integer
reflections $h,k,L+\qr$  have $\kv=\gv+\pv$ for $\pv=\hf.0,\qr$ and  $\kv=\gv-\pv$ for $\pv=-\hf.0,-\qr$
the reverse is true for reflections with $l=L+\qr$. If the population of these two domains are $p^+$ and $p^-$
and $\eta=(p^+-p^-)/(p^++p^-)$

 the asymmetries for a set of orthorhombic equivalent reflections measured with the [100] orientation will be
\[\begin{array}{lllll}
 A(hkl)                = \msp\eta M_{yz}+ C_{Nx}(hkl)N_x&\qquad&
A(hk\bar l)           = -\eta M_{yz}  + C_{Nx}(hk\bar l)N_x\\
 A(h\bar kl)           = \msp\eta M_{yz} - C_{Nx}(h\bar kl)N_x&&
A( h\bar k \bar l)    =   -\eta M_{yz}  - C_{Nx}(h\bar k\bar l)N_x\\
\end{array}\]
\[\begin{array}{lccccc}
 \mbox{since}&A(hkl)=-A(hk\bar l)&\mbox{\&}&A(h\bar kl)=-A(h\bar k\bar l) &C_{Nx}(hkl) =-C_{Nx}(hk\bar l)=1\\
 \mbox{but}&A(h\bar kl)\ne A(hkl)&\mbox{so}&&C_{Nx}(h\bar kl) =-C_{_Nx}(hkl)=-1
\end{array}\]
With these relationships the asymmetries measured with the [100] axis orientation can be used to obtain
values for the quantities $\eta M_{yz}$ and $C_{Nx}N_x$ for each set of reflections measured.
If
\begin{eqnarray*}\langle {\cal A}(hkl)\rangle=&\frac1n\left(\sum_{k \ge 0,l>0} A(hkl) -\sum_{k \ge 0,l<0} A(hkl)\right)\quad \mbox{and}\\
 \langle {\cal A}(h\bar kl)\rangle=&\frac1n\left(\sum_{k \le 0,l<0} A(h k l) -\sum_{k \le 0,l>0} A(hkl)\right)\\
 & \mbox{ where $n$ is the number of terms in each, then}\\
 \end{eqnarray*}
\[  \eta M_{yz}=\left( \langle {\cal A}(hkl)\rangle-\langle {\cal A}(h\bar kl)\rangle\right)/2\quad\mbox{and }\quad
 N_x= \left(\langle {\cal A}(hkl)\rangle+\langle {\cal A}(h\bar kl)\rangle\right)/2\]
 The values obtained from the data measured at 25~K in 5~T are given in table~~\ref{sumdif}. The value of
 $\eta$ estimated by scaling the values of $M_{yz}$ calculated using the structure \cite{kim:08} to the observations was -0.4(3). This result suggests that even a field as low as 1~T is sufficient to favour one out of the two chiral domains but that this initial inbalance 
is not increased further by application of a magnetic field alone. The large standard deviation in $\eta$ shows that the fit with the calculations is 
not very good and for several reflections even the relative signs are inconsistent.
\begin{table}[htdp]
\caption{The averages  $\langle{\cal A}(hkl)\rangle$ and $\langle{\cal A}(h\bar kl)\rangle$ of the polarised neutron intensity asymmetries and the values of 
$\eta M_{yz}$ and $C_xN_x$ deduced from the measurements on \ymno\ at 25~K with 5~T applied parallel to $[100]$. The calculated values were obtained in
least squares fits which gave $\eta=-0.4(3)$ and the Mn$^{4+}$ displacement vector $\delv=-0.042(8),-0.009(8),-0.004(8)$.}
\begin{center}{\footnotesize\setlength{\hw}{-1.2ex}
\begin{tabular}{lllllll}
\hline
&&&\multicolumn{2}{c}{$\eta M_{yz}$}&\multicolumn{2}{c}{$N_x$}\\[\hw]
\multicolumn{1}{c}{$h\ k\ l$}&
\multicolumn{1}{c}{$\langle{\cal A}(hkl)\rangle$}&\multicolumn{1}{c}{$\langle{\cal A}(h\bar kl)\rangle$}\\[\hw]
&&&\multicolumn{1}{c}{Obs}&\multicolumn{1}{c}{Calc}&\multicolumn{1}{c}{Obs}&\multicolumn{1}{c}{Calc}\\
\hline

 0.5 1 0.25&$       -0.034(2)$&$       -0.065(7)$&$       -0.050(4)$&$     -0.0707$&$\msp    0.016(4)$&$\msp   0.0058$\\
 0.5 2 0.25&$       -0.026(4)$&$       -0.047(8)$&$       -0.037(4)$&$     -0.0259$&$\msp    0.010(4)$&$\msp   0.0054$\\
 0.5 3 0.25&$     -0.0111(14)$&$       -0.020(5)$&$       -0.015(2)$&$     -0.0067$&$\msp    0.004(2)$&$\msp   0.0090$\\
 0.5 1 0.75&$\msp  0.0152(12)$&$\msp    0.022(2)$&$\msp  0.0187(14)$&$     -0.0256$&$     -0.0035(14)$&$\msp   0.0086$\\
 0.5 1 1.25&$\msp  0.0208(12)$&$\msp    0.028(4)$&$\msp    0.024(2)$&$\msp  0.0053$&$       -0.003(2)$&$\msp   0.0059$\\
 0.5 2 0.75&$     -0.0074(10)$&$       -0.011(2)$&$     -0.0090(13)$&$\msp  0.0431$&$\msp  0.0016(13)$&$\msp   0.0111$\\
 0.5 2 1.25&$     -0.0128(13)$&$       -0.023(5)$&$       -0.018(2)$&$\msp  0.0383$&$\msp    0.005(2)$&$\msp   0.0050$\\
 0.5 3 0.75&$\msp    0.010(4)$&$\msp    0.009(4)$&$\msp    0.009(3)$&$\msp  0.0276$&$\msp    0.001(3)$&$\msp   0.0179$\\
 0.5 1 1.75&$     -0.0107(11)$&$       -0.015(2)$&$     -0.0127(11)$&$     -0.0061$&$\msp  0.0020(11)$&$      -0.0003$\\
 0.5 1 2.25&$\msp    0.005(2)$&$\msp    0.004(2)$&$\msp  0.0044(11)$&$\msp  0.0082$&$\msp  0.0002(11)$&$      -0.0032$\\
 0.5 2 1.75&$       -0.014(2)$&$       -0.018(4)$&$       -0.016(2)$&$\msp  0.0022$&$\msp    0.002(2)$&$\msp   0.0040$\\
 0.5 3 1.75&$      -0.016(10)$&$      -0.027(10)$&$       -0.022(7)$&$     -0.0083$&$\msp    0.006(7)$&$      -0.0113$\\
 \hline
\end{tabular}
}\end{center}
\label{sumdif}
\end{table}
The poor fit is not altogether surprising as the  
 diffraction intensities and the polarisation analysis results for $h0l$ reflections are largely dependent on the dominant $x$ 
 component of the magnetic moments. Whereas $M_{yz}$ depends on the $y$ and $z$ components of the interaction vectors 
 and on the phase difference between them to which the unpolarised neutron intensity and even the polarisation analysis of $h0l$ 
 reflections  are insensitive.  On the other hand $N_x$  depends on the magnetic structure parameters mainly
 through  the $x$ component of the interaction vector and is relatively insensitive to the $y$ and $z$ components of the magnetic 
 moments.  It may therefore still be possible use the values $N_x$ to estimate the size of the atomic displacements leading to 
 nuclear scattering in the magnetic reflections. $N_y$ and $\eta M_{zx}$ can not be extracted separately from the [010] axis 
 data since only reflections $hkl$ with positive $k$ could be measured.
 \p
\begin{table}[htdp]
\caption{Relationships between the phase differences $\phi_x-\phi_N$ for orthorhombically related reflections; (a) With the 
symmetry of the magnetic structure, (b) with reversal of the displacement 
direction relative to the moment direction for atoms related by $m_z$.}
\begin{center}
\setlength{\hw}{-1.5ex}
\begin{tabular}{ccccc}
\hline
Reflection&\multicolumn{1}{c}{$\phi_N-\phi_x$}&$C_{Nx}\ ^a$&\multicolumn{1}{c}{$\phi_n-\phi_c$}&$C_{Nx}\ ^b$\\
\hline
$\ h\  k\  l$          &$\phi_{Nx}      $&$+1$&$\phi_{Nx}       $&$+1$\\
$\ h\   k\ \bar l$     &$\phi_{Nx}+2n\pi $&$+1$&$\phi_{Nx}+(2n+1)\pi   $&$-1$\\
$\ \bar h\ k\ l$       &$\phi_{Nx}+(2n+1)\pi  $&$-1$&$\phi_{Nx}+2n\pi  $&$+1$\\
$\ \bar h\  k\ \bar l$ &$\phi_{Nx}+(2n+1)\pi  $&$-1$&$\phi_{Nx}+(2n+1)  $&$-1$\\
$\ h\ \bar k\ l$       &$-\phi_{Nx}+(2n+1)\pi $&$-1$&$-\phi_{Nx} +(2n+1)$&$-1$\\
\hline
\end{tabular}\\[0.5ex]
\end{center}
\label{symdisp}
\end{table}%
 The relationships between the $\phi_N$ of  orthorhombically related reflections depend on those between the displacements of
atoms related by the equivalent symmetry operation. In a model in which  $A_N$ is due
solely to the displacements of magnetic atoms and that the relationships between the directions of these displacements for the 
different magnetic  sub-lattices are the same as those relating the corresponding magnetic moments then the phase differences 
$\phi_x-\phi_N$ are as shown in column 3 of \tabl{symdisp}. To obtain the observed
symmetry $C_{Nx}(hkl)=-C_{Nx}(hk\bar l)=C_{Nx}(\bar hkl)=-C_{Nx}(\bar hk\bar l)$  the relative signs of the displacements of pairs of atoms 
related by the mirror plane $m_z$ must be reversed with respect to those of their magnetic moments. With this symmetry however 
the displacements on the Mn$^{3+}$ sites which lie on this mirror plane are zero.
 From equation \ref{fcal} the appropriate nuclear structure factor
 is
\begin{eqnarray*}
&N_\tau =& \imath\sum_{i}b_i\kv\cdot\delta_{io}\exp(\kv\cdot\rv_i\pm \phi_i)/2
\quad\mbox{and for the [100] axis data}\\ 
 &N_x=&2\Re (N_\pv^*M_{\perp x})/|\Qv|^2 
\end{eqnarray*}
A least squares fit of this model, in which the nuclear scattering is due just to displacements of the Mn$^{4+}$ ions, to the values of $N_x$ in \tabl{sumdif} gave $\delv=-0.042(8),-0.009(8),-0.004(8)$, the values of $N_x$ calculated with these displacements are also listed in  \tabl{sumdif}. The fit is not very good which  suggests that the displacements may not  be confined the the magnetic ions; however the results do show that displacements in Mn$^{4+}$ ions of the of the order of 0.1 \AA\ would be required to account for the observations.
\section {Conclusions}
The results  of the present experiments show that useful information about magnetic structure can be obtained from measurements of the
polarisation dependence of the intensities of magnetic satellite reflections of antiferromagnetic chiral crystals. Analysis of the 
relationships between groups of reflections related by the symmetry operations of the paramagnetic space group yield unique information
about the phase relationships between different components of the magnetic interaction vectors and hence the symmetry of the magnetic structure. This gives access to details of the magnetic
structure to which unpolarised neutron intensity measurements are insensitive. The symmetry of the results obtained here confirm that of the proposed structure \cite{kim:08} and suggest that fields as low as 1~T are able to stabilise one out of the two possible chiral domains although it seems that a mono-domain state cannot be achieved by applying magnetic field alone. On the other hand the actual magnitude of the polarisation asymmetries observed  agree rather poorly with those calculated from the parameters of the proposed structure suggesting that this structure may not describe the behaviour  of the $y$ and $z$ components of the magnetic moments correctly. The results also show that when a magnetic field $ \ge$ 1~T is applied parallel to [100] in \ymno\ there is 
an inequivalence between the polarisation asymmetries of Friedel pairs of reflections which suggests a small nuclear scattering contribution to the otherwise purely magnetic reflections. This can be due to small atomic displacements which are modulated with the same wave-vector as the magnetic moments. The symmetry observed in the acentric contribution to the observations is different from that given by
the magnetic moment distribution and constrains the displacements of atoms in the $4h$ (Mn$^{3+}$) sites to be zero.The magnitude of the effect observed 
requires displacements of order 0.1 \AA\ in the Mn$^{4+}$ sites whereas ionic displacements of order only 0.0005 \AA/T induced by an applied magnetic field  have been observed in measurements of TbMnO$_3$ using polarised synchrotron radiation \cite{walker:11}, the dominant displacements being of oxygen. The acentric components of the intensity asymmetry observed in the present experiment on \ymno\ do not
appear to have significant field dependence. The poor agreement between the observations and calculations, based on displacement of manganese atoms only, suggests that in this case also the dominant shifts may be of oxygen. More extensive data are needed to enable
such shifts to be quantified.

\end{document}